\documentclass[10pt, article, twocolumn]{IEEEtran}
\usepackage{amssymb,subfigure,bm,algorithm,algorithmic,stfloats}
\usepackage[dvips]{graphicx}
\usepackage[cmex10]{amsmath}
\usepackage{epsfig,epsf,epstopdf, amssymb,amsfonts,cite,amsmath,epsfig,amsthm,bm,subfigure,fancyhdr,setspace,graphicx,color,mathtools}
\usepackage{epstopdf}
\IEEEoverridecommandlockouts
\newcommand{\bi}{\begin{itemize}}
\newcommand{\ei}{\end{itemize}}
\newcommand{\beq}{\begin{equation}}
\newcommand{\eeq}{\end{equation}}
\newcommand{\bqn}{\begin{eqnarray*}}
\newcommand{\eqn}{\end{eqnarray*}}
\newcommand{\ba}{\begin{array}}
\newcommand{\ea}{\end{array}}
\newcommand{\nn}{\nonumber}
\DeclarePairedDelimiter\abs{\lvert}{\rvert}

\begin{document}

\title{Asynchronous Linear Modulation Classification with Multiple Sensors via Generalized EM Algorithm}

\author{Onur Ozdemir\thanks{$^\dagger$ Boston Fusion Corp.,
1 Van de Graaff Drive Suite 107,
Burlington, MA, USA.  He was with Andro Computational Solutions, Rome, NY when this work was performed. }$^\dagger$,~\IEEEmembership{Member,~IEEE}, Thakshila   Wimalajeewa\thanks{$^{\dagger\dagger}$ Dept. of EECS, Syracuse University, Syracuse, NY, USA, Email: twwewelw@syr.edu, varshney@syr.edu}$^{\dagger\dagger}$ ~\IEEEmembership{Member,~IEEE},   Berkan    Dulek\thanks{$^{\ddagger}$ Department of Electrical and Electronics Engineering, Hacettepe University, Beytepe Campus, Ankara, Turkey}$^{\ddagger}$,  ~\IEEEmembership{Member,~IEEE},  Pramod   K. Varshney$^{\dagger\dagger}$~\IEEEmembership{Fellow,~IEEE} and Wei   Su\thanks{$^{\ddagger\dagger}$ Army CERDEC,
Aberdeen Proving Ground, MD, USA}$^{\ddagger\dagger}$~\IEEEmembership{Fellow,~IEEE}}


\maketitle
\begin{abstract}
In this paper, we consider the problem of automatic modulation classification with multiple sensors  in the presence of unknown time offset, phase offset and received signal amplitude. We develop a novel hybrid maximum likelihood (HML) classification scheme  based on a generalized expectation maximization (GEM) algorithm. GEM is capable of finding  ML estimates numerically that  are extremely hard to obtain otherwise. Assuming  a  good initialization technique  is available for GEM, we show that the classification performance can be greatly improved  with multiple sensors  compared to that with a single sensor, especially when the signal-to-noise ratio (SNR) is low.   We further demonstrate   the superior  performance of our approach when   simulated annealing (SA) with uniform as well as nonuniform grids   is employed for  initialization of GEM  in low SNR regions.
The proposed GEM based  approach employs only a small number of samples (in the order of hundreds) at a given sensor node to perform both time and phase synchronization, signal power estimation, followed by modulation classification.   We provide simulation results to show  the computational  efficiency and effectiveness of the proposed algorithm.
\end{abstract}
\footnotetext[1]{This material is in part based upon work supported by the US Government. US Government Distribution Statement A: Public Release Unlimited. }


\section{Introduction}\label{sec:int}
The problem of automatic modulation classification (AMC) is concerned with determining the modulation type of a received noisy communication signal. With the recent advances in software defined radios, AMC is becoming an integral part of various cognitive radio applications that use adaptive modulation techniques, e.g., adaptive cognitive radios for space communications \cite{hamkins_06}. Widely used  AMC methods can be divided into two general classes: i) likelihood-based (LB)   and ii) feature-based (FB) methods. An extensive overview of these methods is given in  \cite{su_iet07}.
The LB method is based on the likelihood function of the received signal, where the decision is made using a Bayesian hypothesis testing framework. A classifier obtained by the LB method is optimal in the Bayesian sense, i.e., it minimizes the probability of classification error.  Computation of  the likelihood function is challenging when  there are unknown parameters. Various LB based AMC techniques have been proposed in the literature depending on how the unknown parameters are treated. These techniques are known  as generalized likelihood ratio test (GLRT), average likelihood ratio test (ALRT) and hybrid likelihood ratio test (HLRT) \cite{su_tsmc_11}. Despite its computational appeal/lower computational complexity, the traditional GLRT \cite{poor_det} has been shown to provide poor performance in classifying nested constellation schemes such as QAM \cite{su_sarnoff05}.
In ALRT \cite{su_tsmc_11}, which is a fully Bayesian approach, the conditional likelihood function  is averaged over the  unknown signal parameters by assuming certain prior distributions, thereby converting the problem into a simple hypothesis testing problem. In the HLRT approach \cite{su_tsmc_11}, the likelihood function is marginalized over the unknown constellation symbols and then the resulting average likelihood function is maximized over the remaining parameters which are treated as deterministic  unknowns.
A variant of HLRT is quasi HLRT (qHLRT) \cite{dobre_twc09,silva_tc11}, where the unknown signal parameters  are replaced by  their moment-based estimates.


A large number of   AMC techniques developed so far  make the common assumption that perfect timing information is available at the receiver. This assumption is unrealistic for practical AMC scenarios due to a number  of reasons. First, AMC usually needs to be performed in a noncooperative environment. Therefore, there is no training sequence available at the receiver for accurate time synchronization. The receiver has to employ blind time synchronization techniques which result in residual errors, namely time offsets in the received signal \cite{mengali_97}. Second, AMC is generally based on batch techniques where only a finite number of samples are available for classification. Both blind synchronization and modulation classification need to be carried out using these limited number of samples. In other words, the receiver does not have the luxury to obtain massive amounts of data for perfect time synchronization in practice.  This fact should be taken into account in the design of an AMC algorithm.
 Although time offsets are unavoidable in  most AMC scenarios, there have been only a few research works that have addressed this issue \cite{weber_tc98,namazi_tc02,polydoros_95,guan_wcnc08,silva_tc11_2}. Among these works, MFSK modulations are considered in \cite{weber_tc98,namazi_tc02}, whereas PSK and QAM modulations are the focus in this paper. In one of the earlier papers on  likelihood based AMC \cite{polydoros_95}, the authors consider only PSK modulations and derive approximate forms of the likelihood functions that are obtained by marginalizing out both phase and time offsets in addition to unknown constellation symbols (i.e. ALRT) under low signal-to-noise ratio (SNR) assumptions. It has been shown in \cite{polydoros_95} that the general ALRT does not have a closed-form analytical  expression. In \cite{guan_wcnc08}, the authors consider only QAM modulations. They adopt the square timing recovery technique for blind synchronization followed by a cumulant based hierarchical modulation classifier. Aside from its limitation to only QAM modulations, the  method proposed in \cite{guan_wcnc08} requires a large number of samples for accurate blind synchronization and it can only classify a limited number of QAM modulations for which appropriate cumulants need to be selected. More recently, the authors in \cite{silva_tc11_2} consider linear modulations (amplitude-phase modulations). They propose moment-based estimators to estimate the unknown time offset along with unknown phase offset and SNR in their  qHLRT approach. They collect a number of samples for estimation and then an additional number of samples for modulation classification. The shortcomings of this approach are two-fold. First, moment-based estimators do not always provide meaningful estimates, especially when the number of samples for estimation is small. For example in \cite{silva_tc11_2}, $10000$ samples are used to obtain acceptable classification performance. Second, moment-based estimators do not necessarily maximize the likelihood function which results in unavoidable sub-optimality in the classification step. The reason behind the use of  moment-based estimators for AMC is the computational complexity associated with maximum likelihood (ML) estimators as pointed out in \cite{silva_tc11_2}.


In this paper, we focus on the problem of classifying   linear modulations in the presence of unknown time offset, phase offset and signal amplitude with multiple sensors. This work is based on our initial work for asynchronous  AMC with a single sensor \cite{ozdemir_globecom13}. In \cite{ozdemir_globecom13}, we developed a hybrid maximum likelihood (HML)\footnote{HML is also referred to as the hybrid likelihood ratio test (HLRT) in some  modulation classification literature \cite{su_iet07,dobre_twc09,silva_tc11_2,silva_tc11}.} based approach to AMC in the presence of unknown time offset, phase offset and signal amplitude with a single sensor. To find unknown parameters via maximum likelihood estimation, a computationally efficient numerical algorithm was proposed based on generalized expectation maximization (GEM). The GEM algorithm provides a tractable procedure to obtain ML estimates which are extremely hard to obtain otherwise.  In the current work, we aim to improve the performance of HML  based AMC in the low SNR region by employing multiple sensors. Solution to this problem cannot be obtained as a straightforward extension of the approach developed for the single-sensor case as discussed next. We assume that the observations collected at multiple sensors are available at a fusion center  to perform classification.    In the multiple sensor case, multiple nodes observe  the same signal whose  constellations are assumed to be randomly distributed, thus,  the observations collected at multiple nodes are not independent.   The unknowns related to each node are  estimated based on the likelihood function computed using the joint probability density function (pdf). Finding the global maximum of the joint likelihood function with respect to unknowns is a high  dimensional non-convex optimization problem which is prohibitively complex to solve in general. Further,  there is coupling
between the unknowns of different sensors due to common
unknown constellation symbols, i.e., the problem cannot be
decoupled across sensors into multiple lower dimensional optimization problems.  However, the GEM-based algorithm   developed in this  paper decouples
this problem by computing the posterior expectations of the constellation symbols and
uses them for decoupled estimation (across sensors) at each iteration. Computation of these posterior
expectations exploits this coupling effect, i.e., jointly use the information from all the radios.
Therefore, it is the crucial step to enable data fusion. Further,  we observe that the performance of the GEM based approach for joint classification  is more susceptible  to parameter initialization of the GEM algorithm compared to that with the  single sensor case. Provided that a good initialization technique is available, it is shown that the  GEM algorithm  with multiple sensors provides  promising results irrespective of the nature of the other relevant parameters such as the channel SNR, the number of sensors and the true modulation format.

After deriving the GEM algorithm for modulation classification with multiple sensors, we first evaluate the performance assuming that an initialization technique is given. For any initialization technique, we can express the initial values as true values of the unknown  parameters plus some error. The error term determines how good the initialization technique is. Under  this assumption, we provide simulation results to show the performance gain achievable with multiple sensors in the presence of unknown time offset, phase offset and channel gain, and also illustrate  the impact of the GEM initialization on the overall performance. It is seen that, when the initialization error is sufficiently small, the proposed GEM algorithm provides closer   performance to the Clairvoyant classifier \cite{mendel_tc_00} (which assumes that the unknown parameters are known)  with multiple sensors.  Next, we consider simulated  annealing (SA) based stochastic initialization technique for GEM initialization  which is shown to provide good results in the  low SNR region with multiple sensors. We further  investigate the trade-off between the number of samples to be collected by each node versus the number of nodes, and the performance of  AMC with GEM as the number of unknowns is varied.  

 Our approach is  applicable to all QAM and PSK modulations especially in the low  SNR region.  Moreover, the proposed scheme  employs only a small number of samples (in the order of hundreds) at a given node, as opposed to thousands as in  \cite{silva_tc11_2}, to perform both time synchronization and modulation classification. More importantly, since the proposed approach maximizes the original likelihood function, it is expected to perform better than the qHLRT approach. The proposed approach also enables maximum a posteriori (MAP) decoding of the unknown constellation symbol sequence as a by-product of the GEM algorithm. Our simulation results show that the proposed approach provides excellent classification performance with, for example, only $N=100$ samples per node for a modulation classification scenario involving 8-PSK, 8-QAM, 16-PSK, and 16-QAM modulation formats.

The rest of the paper is organized as follows. In Section \ref{sec:prob}, we introduce the system model and formulate the HML based modulation classification problem with multiple sensors. The details of our proposed GEM based classifier are presented in Section \ref{sec:em} and subsections therein. We provide numerical results to depict the performance of the proposed approach in Section \ref{sec:num}. Finally, concluding remarks along with avenues for future work are provided in Section \ref{sec:conc}.


%

\section{Problem Formulation}\label{sec:prob}
We consider $L$ radio receivers observing a linearly modulated communication signal that undergoes block fading. The received baseband signal at the $l$-th radio can be expressed as
\beq
y_l(t) = a_l  e^{j\theta_l} \sum_{n} I_n g(t-nT-\varepsilon_l T) + w_l(t) , \quad 0\leq t \leq T_0\label{eq:simp_rec}
\eeq
for $l=1,\cdots, L$ where $T_0$ is the observation interval, $T$ is the symbol duration, $g(t)$ is the transmitted pulse, $I_n$ is the $n^{th}$ complex constellation of the transmitted symbol, $w_l(t)$ is the additive complex zero-mean white Gaussian noise process with two-sided power spectral density (PSD) $N_0/2$, $a_l > 0$ is the channel gain, $\theta_l\in [-\pi,\pi)$ is the channel phase\footnote{The phase term $\theta_l$ subsumes both the channel phase and the residual phase offset at the receiver.}, and $\varepsilon_l T$ is the residual time offset at the receiver. We assume that the estimation of the pulse shape $g(\cdot)$, the symbol duration $T$ and the carrier frequency has been accomplished at each  receiver. These are commonly made assumptions in the modulation classification literature \cite{su_iet07,sadler_tc00,guan_wcnc08,dobre_twc09,cabric_cl11,silva_tc11_2,silva_tc11}, and these estimates can be obtained using the techniques outlined in \cite{sadler_tc00}. Without loss of generality, we also assume $\varepsilon_l\in[0,1)$. In this model, $\{a_l,\theta_l,\epsilon_l\}_{l=1}^L$ for $l=1, \cdots, L$ and $\{I_n\}_{n=0}^{N-1} $  are the unknown signal parameters. Suppose there are $S$ candidate modulation formats under consideration and let  $I_n^{(i)}$ denote the $n^{th}$ constellation symbol corresponding to modulation $i\in\{1,\ldots,S\}$. The goal is to identify the correct modulation format from $S$ candidate formats based on the observation $\{y_l(t)\}_{l=1}^L$.

Let $\mathcal H_i$ denote the hypothesis associated with the modulation $i$. In a Bayesian setting, the optimal classifier which  minimizes the  probability of classification error is the MAP classifier. We assume that each modulation is equally likely, i.e., each has the same prior probability. In this case, the optimal classifier takes the form of a ML classifier. As mentioned in Section \ref{sec:int}, we focus on the HML approach \cite{su_iet07}, where the likelihood function is marginalized over the unknown random  constellation symbols $I_n$ and then maximized over the remaining unknown (nuisance) parameters. Let $\mathbf{u}\triangleq \left[a_1,\ldots,a_L,\theta_1,\ldots,\theta_L,\varepsilon_1,\ldots,\varepsilon_L\right]$ denote the deterministic  unknown parameter vector. We define $s_l(t)$ as
\beq
s_l(t)\triangleq a_l  e^{j\theta_l} \sum_{n} I_n g(t-nT-\varepsilon_l T), ~ 0\leq t \leq T_0
\eeq
for $l=1,\cdots, L$.
Let $\mathbf{y}_l$ denote a vector representation of $y_l(t)$. We also define $\mathbf{I} \triangleq [I_0,\ldots,I_{N-1}]^T$ and $\mathbf{y}\triangleq[\mathbf{y}_1^T,\ldots,\mathbf{y}_L^T]^T$ where $(\cdot)^T$ denotes vector/matrix transpose. It should be clear from the context when $T$ represents symbol duration or transpose. It can be shown that the conditional likelihood function of the noisy received signals is given by \cite{poor_det}
\begin{align}
&p_i(\mathbf{y}|\mathbf{u},\mathbf{I})\propto  \label{eq:llfc} \\
&\exp\left\{\frac{2}{N_0} \sum_{l=1}^L\int_0^{T_0}\Re \left\{y_l(t)s_l^*(t)\right\} dt  - \frac{1}{N_0}\sum_{l=1}^L\int_0^{T_0}\abs{s_l(t)}^2 dt\right\} \nn
\end{align}
where $p_i(\cdot) \triangleq p(\cdot|\mathcal  H_i)$ and $\Re(\cdot)$ denotes the real part of a complex number. The observation interval $T_0$ is based on  designer's choice, so we assume that $T_0$ is a multiple of $T$ and define $N\triangleq T_0/T$. The symbol pulse $g(t)$ is a finite-length pulse (e.g., symmetrically truncated root-raised cosine (RRC) pulse) with duration $T_p$. We assume that $T_0\gg T_p$, i.e., the observation interval is much larger than the duration of the transmit pulse. This assumption is well justified in all practical modulation classification applications as it is typical to observe at least $\sim100$ symbols before making a decision. Under  these assumptions, we get the following two expressions \cite{mengali_97}
\begin{align}
\int_0^{T_0}&\Re\{y_l(t)s^*(t)\} dt =  \\
& a_l \Re\left\{ e^{-j\theta_l}\sum_{n=0}^{N-1} I_n^* \int_0^{T_0} y_l(t) g^*(t-nT-\varepsilon_l T) dt \right\}\nn
\end{align}
\beq
\int_0^{T_0}\abs{s_l(t)}^2 dt \approx E_g a_l^2 \sum_{n=0}^{N-1} \abs{I_n}^2 \label{eq:approx2}
\eeq
where $E_g$ is the energy of the transmit pulse
\beq
E_g \triangleq \int_{-\infty}^{\infty}g^2(t) dt.
\eeq
The approximation in \eqref{eq:approx2} is based on the assumption that $T_0\gg T_p$. In other words, the contribution of the symbols in the beginning and at the end of the observation interval to the total energy of the received signal will be negligible for $T_0\gg T_p$. With this approximation, the likelihood function can be written as
\begin{align}
& p_i(\mathbf{y}|\mathbf{u},\mathbf{I}) \propto \nn \\
&\exp\left\{\frac{2}{N_0} \sum_{n=0}^{N-1} \sum_{l=1}^La_l\Re\left\{I_n^* e^{-j\theta_l} \int_0^{T_0}y_l(t) g^*(t-nT-\varepsilon_l T) dt \right\} \right\} \nn \\
&\cdot \exp\left\{-\frac{E_g }{N_0} \sum_{n=0}^{N-1}\abs{I_n}^2 \sum_{l=1}^L a_l^2 \right\}. \label{eq:llfc2}
\end{align}
Note that \eqref{eq:llfc2} now denotes approximate proportionality due to  \eqref{eq:approx2}. Now we turn our attention to our original problem where we need to marginalize the distribution over the constellation symbols, i.e., we need to compute
\beq
p_i(\mathbf{y}|\mathbf{u}) = \sum_{\mathbf{I}^{(i)}} p_i(\mathbf{y}|\mathbf{u},\mathbf{I}^{(i)}) P(\mathbf{I}^{(i)}) \label{eq:1}
\eeq
where $\mathbf{I}^{(i)} \triangleq [I_0^{(i)},\ldots,I_{N-1}^{(i)}]^T$ denotes the received constellation symbol vector under hypothesis $\mathcal H_i$. 
The above expression can be simplified by noting that the constellation symbols are independent and identically distributed (i.i.d.) with $P(I_n^{(i)})=1/M_i$, where $M_i$ is the cardinality\footnote{For example, $M_i=2$ for BPSK and $M_i=16$ for 16-QAM.}  of the constellation symbol set  for hypothesis $\mathcal H_i$. The resulting log-likelihood function $\Lambda_i(\mathbf{u})\triangleq \ln p_i(\mathbf{y}|\mathbf{u})$ is given in  \eqref{eq:lf1}.
\begin{figure*}
\begin{eqnarray}
\Lambda_i(\mathbf{u}) &= &\sum_{n=0}^{N-1} \ln \left(\sum_{k=1}^{M_i} \exp\left\{\frac{2}{N_0}\sum_{l=1}^L a_l\text{Re}\left\{I_n^{k*} e^{-j\theta_l} \int_0^{T_0}y_l(t) g^*(t-nT-\varepsilon_l T) dt \right\} - \frac{E_g }{N_0}\sum_{l=1}^L a_l^2\abs{I_n^{k}}^2\right\}\right) \nn\\
&-&N\ln M_i\label{eq:lf1}
\end{eqnarray}
\underline{\hspace{\textwidth}}\vspace{-.15cm}
\end{figure*}
The  maximum likelihood estimate (MLE) of $\mathbf{u}$ under $\mathcal H_i$ is given as
\beq
\hat{\mathbf{u}}_i = \arg\max_{\mathbf{u}} \Lambda_i(\mathbf{u}). \label{mle1}
\eeq
Finally, the HML modulation classifier is
\beq
\hat{i} = \arg\max_{i} \Lambda_i(\hat{\mathbf{u}}_i).  \label{hml}
\eeq
It is noted that in (\ref{hml}), $\Lambda_i(\hat{\mathbf{u}}_i)$ is computed for each $i$ based on (\ref{eq:lf1}) for $i=1, \cdots, S$ and the index corresponding to the maximum is selected.
Due to the marginalization over the constellation symbols, the resulting ML estimation problem in \eqref{mle1} is not tractable. This is because it is a $3 \times L$ dimensional non-convex optimization problem and there is no closed-form solution. Therefore, finding the MLE $\hat{\mathbf{u}}_i$ from \eqref{mle1} would normally require an exhaustive search which is computationally expensive and is impractical in real AMC applications. In order to solve this problem, we propose an efficient algorithm in the next section which is based on the Generalized Expectation Maximization (GEM) algorithm \cite{rubin_jrs77}.

\section{The EM Algorithm for AMC}\label{sec:em}
The Expectation Maximization (EM) algorithm is an iterative method which enables the computation of ML estimates. It is especially well suited to problems where ML estimation is intractable due to the presence of hidden (unobserved) data. For the  problem addressed in this paper, the actual sequence of transmitted constellation symbols $\mathbf{I}$ can be treated as  hidden data. We can formally describe the EM algorithm for our problem in (\ref{eq:1}) as follows \cite{rubin_jrs77}. Let us define the so-called \emph{complete data} $\mathbf{x}=[\mathbf{y}^T,\mathbf{I}^T]^T$. Starting from an initial estimate $\hat{\mathbf{u}}_i^{(0)}$ under the hypothesis $\mathcal H_i$, the EM algorithm performs the following two steps: the expectation step (E-step) and the maximization step (M-step).
\begin{align}
&\text{\textbf{E-step:}}\quad Q(\mathbf{u}|\hat{\mathbf{u}}_i^{(r)})=E\left\{\ln p_i(\mathbf{x}|\mathbf{u})|\mathbf{y},\hat{\mathbf{u}}_i^{(r)}\right\},\label{eq:estep}\\
&\text{\textbf{M-step:}}\quad \hat{\mathbf{u}}_i^{(r+1)}=\arg\max_{\mathbf{u}} Q(\mathbf{u}|\hat{\mathbf{u}}_i^{(r)}).\label{eq:mstep}
\end{align}
Given the fact that the unknown parameter vector $\mathbf{u}$ is independent of the transmitted constellation symbols $\mathbf{I}$, the E-step in (\ref{eq:estep}) reduces to
\beq
Q(\mathbf{u}|\hat{\mathbf{u}}_i^{(r)}) = \sum_{\mathbf{I}} \ln p_i(\mathbf{y}|\mathbf{I},\mathbf{u}) P_i\left(\mathbf{I}|\mathbf{y},\hat{\mathbf{u}}_i^{(r)}\right), \label{eq:estep2}
\eeq
where $\ln p_i(\mathbf{y}|\mathbf{I},\mathbf{u})$ is as given in \eqref{eq:llfc2}. Suppose we have $\hat{\mathbf{u}}_i^{(r)}$ at the end of the $r$-th iteration. We define $y_{n,l}^{(r)}$ as
\beq
y_{n,l}^{(r)} \triangleq y_l(nT+\hat{\varepsilon_l}^{(r)}T) = \int_0^{T_0}y_l(t) g^*(t-nT-\hat{\varepsilon_l}^{(r)}T) dt. \label{eq:y_sample}
\eeq
Let  $\alpha_n^{m,(r)}\triangleq P_i\left(I_n=I^{m}|y_{n,1}^{(r)}, \cdots, y_{n,L}^{(r)},\hat{\mathbf{u}}_i^{(r)}\right)$, $m=1,\ldots,M_i$, denote the \emph{a posteriori} probability of the $n^{th}$ unknown constellation symbol which can be calculated as
\begin{eqnarray}
\alpha_n^{m,(r)} &\triangleq& P_i\left(I_n=I^{m}|y_{n,1}^{(r)}, \cdots, y_{n,L}^{(r)},\hat{\mathbf{u}}_i^{(r)}\right) \\
&=& \frac{p_i\left(I_n=I^{m}, y_{n,1}^{(r)}, \cdots, y_{n,L}^{(r)}|\hat{\mathbf{u}}_i^{(r)}\right)}{P_i\left(y_{n,1}^{(r)}, \cdots, y_{n,L}^{(r)}|\hat{\mathbf{u}}_i^{(r)}\right)}\nonumber\\
&\stackrel{(a)}{=}&\frac{p_i\left(y_{n,1}^{(r)}, \cdots, y_{n,L}^{(r)}|I_n=I^{m},\hat{\mathbf{u}}_i^{(r)}\right)}{\sum\limits_{k=1}^{M_i} p_i\left(y_{n,1}^{(r)}, \cdots, y_{n,L}^{(r)}|I_n=I^{k},\hat{\mathbf{u}}_i^{(r)}\right)}  \nonumber \\
& = &\frac{\exp\left(-\sum\limits_{l=1}^L\abs{y_{n,l}^{(r)}- \hat{a}_l^{(r)} e^{j \hat{\theta_l}^{(r)}} I^m}^2 / N_0\right)}
{\sum\limits_{k=1}^{M_i} \exp\left(-\sum\limits_{l=1}^L\abs{y_{n,l}^{(r)}- \hat{a}_l^{(r)} e^{j \hat{\theta_l}^{(r)}} I^k}^2 / N_0\right)}.\nonumber
\label{eq:post}
\end{eqnarray}
In $(a)$, we have used the assumption that each data symbol has the same \emph{a priori} probability, i.e., $P_i\left(I_n=I^{m}|\hat{\mathbf{u}}_i^{(r)}\right)=1/M_i$, $m=1,\ldots,M_i$. Substituting \eqref{eq:llfc2} in \eqref{eq:estep2} along with $\alpha_n^{m,(r)}$, $Q(\mathbf{u}|\hat{\mathbf{u}}_i^{(r)})$ reduces to
\begin{eqnarray}
Q(\mathbf{u}|\hat{\mathbf{u}}_i^{(r)}) = \sum_{l=1}^L Q_l(\mathbf{u}_{l}|\hat{\mathbf{u}}_{l,i}^{(r)})\label{Q_u}
\end{eqnarray}
where $\mathbf  u_l $ contains the unknowns in  $\mathbf u$ that correspond to the $l$-th node, i.e., $\mathbf  u_l=[a_l , \theta_l, \varepsilon_l]^T$ for $l=1,\cdots, L$ and  $Q_l(\mathbf{u}_{l}|\hat{\mathbf{u}}_{l,i}^{(r)})$ is given by (\ref{Q_l_1}) on the next page.
\begin{figure*}
\begin{eqnarray}
Q_l(\mathbf{u}_{l}|\hat{\mathbf{u}}_{l,i}^{(r)})
 = \frac{2a_l}{N_0} \sum_{n=0}^{N-1} \sum_{m=1}^{M_i} \alpha_n^{m,(r)}  \Re\left\{I_n^{m*} e^{-j\theta_l} \int_0^{T_0}y_l(t) g^*(t-nT-\varepsilon_l T) dt \right\} -
\frac{E_g a_l^2}{N_0} \sum_{n=0}^{N-1} \sum_{m=1}^{M_i} \alpha_n^{m,(r)} \abs{I_n^m}^2. \label{Q_l_1}
\end{eqnarray}
\end{figure*}
Defining   the posterior expectations of the $n^{th}$ transmitted symbol and the average normalized signal energy to be
\beq
\hat{I}_n^{(r)} \triangleq \sum_{m=1}^{M_i} \alpha_n^{m,(r)} I_n^{m},\quad\quad \hat{E}_I^{(r)} \triangleq \sum_{n=0}^{N-1}\sum_{m=1}^{M_i}\alpha_n^{m,(r)} \abs{I_n^m}^2, \label{eq:post2}
\eeq
respectively, (\ref{Q_l_1}) reduces to
\begin{align}
Q_l&(\mathbf{u}_{l}|\hat{\mathbf{u}}_{l,i}^{(r)})= \label{Q_l}\\
&\frac{2 a_l}{N_0} \sum_{n=0}^{N-1} \Re\left\{\hat I_n^{(r)*} e^{-j \theta_l} \int_0^{T_0} y_l(t)g^*(t-nT -\varepsilon_lT) dt\right\} \nonumber\\
&- \frac{E_ga_l^2}{N_0}\hat E^{(r)}_I. \nn
\end{align}
Then the maximization step in (\ref{eq:mstep}) at $r$-the iteration of the EM algorithm reduces to
\begin{eqnarray}
\hat{\mathbf{u}}_{l,i}^{(r+1)} = \underset{\mathbf u_l}{\arg\max} ~ Q_l(\mathbf{u}_{l}|\hat{\mathbf{u}}_{l,i}^{(r)})\label{max_Ql}
\end{eqnarray}
for $l=1, \cdots, L$ where $Q_l(\mathbf{u}_{l}|\hat{\mathbf{u}}_{l,i}^{(r)})$ is as given in (\ref{Q_l}). Note that we now have a three dimensional optimization problem for each sensor as opposed to a single $3\times L$ dimensional original optimization problem. The maximization step in \eqref{max_Ql} can be carried out in two steps:
\begin{align}
&\left(\hat{\theta_l}^{(r+1)},\hat{\varepsilon_l}^{(r+1)}\right)= & \label{mle4}
\\
& \arg\max_{\theta_l,\varepsilon_l} \sum_{n=0}^{N-1}  \Re\left\{\hat{I}_n^{(r)*} e^{-j\theta_l} \int_0^{T_0}y_l(t) g^*(t-nT-\varepsilon_l T) dt \right\}\nn
\end{align}
\begin{align}
& \hat{a}_l^{(r+1)} =\frac{1}{E_g \hat{E}_I^{(r)}}\cdot \label{mle5}\\
& \sum_{n=0}^{N-1}  \Re\left\{\hat{I}_n^{(r)*} e^{-j\hat{\theta_l}^{(r+1)}} \int_0^{T_0}y_l(t) g^*(t-nT-\hat{\varepsilon_l}^{(r+1)} T) dt \right\}\nn
\end{align}
for $l=1,\cdots, L$.
At the $(r+1)$-th iteration, the maximization step in \eqref{mle4} constitutes a weighted correlation (or matched filtering) based estimation of the unknown phase and time offsets. After these estimates are obtained, the estimate of the signal amplitude is computed in closed-form using \eqref{mle5}. Even though the EM algorithm simplifies the ML estimation procedure significantly, the optimization problem in \eqref{mle4} still requires maximization over two dimensions which can be computationally expensive. Thus, in the following, we consider a computationally efficient approach to overcome this issue.

\subsection{Generalized EM}\label{sec_GEM}
The EM algorithm is theoretically guaranteed to converge to a stationary point as long as the  $Q(\mathbf{u}_i|\hat{\mathbf{u}}_i^{(r)})$ function increases at every iteration \cite{wu_ann83}. In other words, the maximization step in \eqref{eq:mstep} can be replaced with an improvement step, which does not impact the convergence property of the EM algorithm. These variants of the EM algorithms are referred to as Generalized EM (GEM) algorithms \cite{rubin_jrs77}. Due to this theoretical result, we can replace the maximization step in \eqref{mle4} with the following `block coordinate ascent' type procedure
\begin{align}
&\hat{\varepsilon_l}^{(r+1)} = \nn\\
&\arg\max_{\varepsilon_l} \sum_{n=0}^{N-1}  \Re\left\{\hat{I}_n^{(r)*} e^{-j \hat{\theta_l}^{(r)}} \int_0^{T_0}y_l(t) g^*(t-nT-\varepsilon_l T) dt \right\},
\label{mle6} \\
& \hat{\theta_l}^{(r+1)} = \tan^{-1}\left(\frac{\Im(\hat{\mathbf{I}}^{(r)^H}\mathbf{y}_l^{(r+1)})}{\Re(\hat{\mathbf{I}}^{(r)^H}\mathbf{y}_l^{(r+1)})}\right), \label{mle7}
\end{align}
for $l=1,\cdots, L$
where $\Im(\cdot)$ denotes the imaginary part of a complex number, $\hat{\mathbf{I}}^{(r)} \triangleq \left[\hat{I}_0^{(r)},\ldots,\hat{I}_{N-1}^{(r)}\right]^T$, and $\mathbf{y}_l^{(r+1)} \triangleq \left[y_{0,l}^{(r+1)},\ldots,y_{N-1,l}^{(r+1)}\right]^T$, in which $y_{n,l}^{(r+1)}$ is obtained from \eqref{eq:y_sample}, i.e., $y_{n,l}^{(r)} \triangleq y_l(nT+\hat{\varepsilon_l}^{(r)}T)$. Note that the above two steps are much simpler to implement than \eqref{mle4}, since \eqref{mle6} requires a line search which can be carried out by methods such as the Newton-Raphson method and \eqref{mle7} is a closed-form expression. Even though the $\left(\hat{\theta_l}^{(r+1)},\hat{\varepsilon_l}^{(r+1)}\right)$ pair obtained by \eqref{mle6}-\eqref{mle7} does not necessarily maximize $Q(\mathbf{u}_i|\hat{\mathbf{u}}_i^{(r)})$, the EM algorithm is guaranteed to converge to a stationary point of the original likelihood function after a sufficient number of iterations.

We should also mention that if the time offset $\varepsilon_l$ is perfectly known, the EM algorithm simplifies  significantly. Under this scenario, the EM algorithm would iterate over \eqref{eq:post2}, \eqref{mle7} and \eqref{mle5} which are all closed-form expressions. Numerical results for such a scenario are provided in Section \ref{sec:num}.

\subsection{Initialization of unknown parameters}\label{EM_Initialization}
The initialization of the EM algorithm, namely  obtaining the initial estimate $\hat{\mathbf{u}}_i^{(0)}$, has a large   impact on the  stationary point the EM will converge to. A good initial point increases the likelihood that the algorithm will converge to the global maximum rather than to some local maxima.  Our GEM based approach for AMC with multiple sensors is seen  susceptible to convergence to a local maxima  as the number of sensors increases unless a good  initialization technique is available.  There are a number of methods that  can be used  to initialize the EM algorithm.  The initial  estimates obtained with any initialization technique can be expressed as the true value of the parameter plus some error. Obviously, larger the error, the worse the initialization technique is. More  specifically, we represent the initial values in the form of  $\hat{\mathbf{u}}_i^{(0)} = {\mathbf{u}} + \boldsymbol\epsilon $ where $\boldsymbol\epsilon$ is a $3L\times 1$ vector which denotes the deviation of the initialization points of unknown parameters from their true values. In Section \ref{initial}, we provide numerical results to illustrate the performance gain achievable by GEM  with multiple sensors as $\boldsymbol\epsilon $ varies. This approach provides insights into how much performance gain is achievable  with proposed GEM approach for AMC with multiple sensors. In the following, we consider a practical scheme for EM initialization, which provides good initial estimates in low SNR regions (which is the most interesting scenario).

\subsubsection{EM Initialization with simulated annealing (SA)}
We adopt a modified SA method which is implemented over a coarse grid and over a predefined finite number of iterations. Specifically, we construct the following grid sets $\Theta \triangleq \{-\pi,-\pi+\Delta_{\theta_l},-\pi+2\Delta_{\theta_l},\ldots,\pi-\Delta_{\theta_l}\}$,
$\xi \triangleq \{0,\Delta_{\varepsilon_l},\ldots,1-\Delta_{\varepsilon_l}\}$, and $A \triangleq \{\Delta_{a_l},2\Delta_{a_l},\ldots,a_{l}^U\}$, where $a_l^U$ is some upper bound which is based on designer's choice and can be selected depending on the channel characteristics for a given scenario.  The increments (denoted by $\Delta$s) determine the resolutions of the grid sets, i.e., how coarse the grids are. Let us define $\Omega \triangleq \Theta \times \xi \times A$. Let $K$ denote the maximum number of iterations and $d$ denote the predefined SA parameter. The parameters $K$ and $d$ are adjusted by the user. The SA algorithm is summarized in Algorithm \ref{SA}. In Algorithm \ref{SA}, a neighbor of a point is defined as one of the points in the one-hop neighborhood of the point.
\begin{algorithm} []
\caption{SA for GEM Initialization under $\mathcal H_i$}
\label{SA}
\begin{algorithmic} [1]
\STATE Randomly  select $\omega_1 \in \Omega$. Initialize $\omega_F = \omega_1$.
\STATE FOR $k=2,\ldots,K$
\STATE $T=d/\log(k)$.
\STATE Randomly select a neighbor of $\omega_k$, denoted by $\omega_N$.
\STATE If $\Lambda_i (\omega_k)\leq \Lambda_i (\omega_N)$, set $\omega_{k+1}=\omega_N$.
\STATE Else, $\omega_{k+1}=\omega_N$ with probability
\beq
\exp\left(\frac{\Lambda_i (\omega_N)-\Lambda_i (\omega_k)}{T\abs{\Lambda_i (\omega_k)}}\right), \nn
\eeq
$\omega_{k+1}=\omega_k$ otherwise.
\STATE If $\Lambda_i (\omega_F)\leq \Lambda_i (\omega_{k+1})$, set $\omega_F = \omega_{k+1}$ and continue.
\STATE ENDFOR
\STATE Set $\hat{\mathbf{u}}_i^{(0)}=\omega_F$.
\end{algorithmic}
\end{algorithm}
 Note that, instead of iterating until convergence, a maximum number of iterations is employed for the SA algorithm. This is in part  to keep the overall computational complexity low and in part due to the fact that the GEM algorithm takes care of the fine maximization step. The  overall goal is to find a `good' initial point for the GEM algorithm. Other methods such as moment-based estimators can also be employed for initialization as long as they have low computational complexity and  provide `good' initial points.


\subsection{GEM Summary}
For clarity, we summarize the proposed GEM based asynchronous modulation classifier (MC) in Algorithm \ref{GEM}. After a classification decision has been made, the MAP decoding of the received symbol sequence can be easily obtained using the final \emph{a posteriori} probabilities $\alpha_n^{m,(r)}$ which have already been calculated in the GEM algorithm for  modulation $\hat{i}$.
\begin{algorithm} []
\caption{GEM Based Asynchronous MC}
\label{GEM}
\begin{algorithmic} [1]
\STATE Set stopping criterion $\delta$.
\STATE FOR $i=1,\ldots,S$
\STATE Set $r=0$. Initialize $\hat{\mathbf{u}}_i^{(0)}$
\STATE For $n = 0,\ldots,N-1$; $m=1,\ldots,M_i$; compute $\alpha_n^{m,(r)}$ from \eqref{eq:post}.
\STATE For $n = 0,\ldots,N-1$; compute $\hat{I}_n^{(r)}$ from \eqref{eq:post2}.
\STATE Compute $\hat{E}_I^{(r)}$ from \eqref{eq:post2}.
\STATE Set $r=r+1$
\STATE Compute $\hat{\varepsilon_l}^{(r+1)}$ using \eqref{mle6}.
\STATE Compute $\hat{\theta_l}^{(r+1)}$ using \eqref{mle7}.
\STATE Compute $\hat{a}^{(r+1)}$ using \eqref{mle5}.
\STATE If $\Lambda_i(\hat{\mathbf{u}}_i^{(r+1)})-\Lambda_i(\hat{\mathbf{u}}_i^{(r)})> \delta$, go to Step $4$, else set
$\hat{\mathbf{u}}_i^ = \hat{\mathbf{u}}_i^{(r+1)}$ and continue.
\STATE ENDFOR
\STATE Final decision $\hat{i} = \arg\max_{i} \Lambda_i(\hat{\mathbf{u}}_i)$.
\end{algorithmic}
\end{algorithm}

\section{Numerical Results}\label{sec:num}
In this section, we provide numerical results to illustrate the performance gain achievable with multiple sensors using   the proposed GEM based modulation classification scheme compared to that with a single sensor. We consider a scenario where $g(t)$ is a symmetrically truncated root-raised-cosine  (RRC) pulse \cite{Proakis:book}, i.e., $g(t)=g(-t)$, with a roll-off factor $\alpha=0.3$ and duration $8T$. Without loss of generality, we assume that $\mathbb{E}\{\abs{I_n}^2\}=1$ and $N_0=1$. The channels between the transmitter and each sensor  are modeled as Rayleigh fading channels, i.e., $a_l$ is a Rayleigh distributed random variable with scale parameter $\sigma$ for $l=1,\cdots, L$. With these assumptions, the channel signal-to-noise ratio (SNR) is $\mathbb{E}\{a_l^2\abs{I_n}^2\}/N_0=2\sigma^2$. We assume $T=1$, $\theta_l \sim \mathcal{U}[-\pi,\pi)$ and $\varepsilon_l \sim \mathcal{U}[0,1)$, for $l=1,\cdots,L$  where $\mathcal{U}[a,b)$ denotes uniform distribution with support $[a,b)$. The observation interval is set as $T_p = NT$. We consider a quaternary classification scenario. The modulations to be classified are 8-PSK, 8-QAM, 16-PSK, and 16-QAM. In the following, we assess the classification performance with respect to different aspects including the channel SNR, the true modulation format,   the initial values of the unknowns used for the  GEM algorithm, number of sensors,  number of samples per node, and the impact of ignoring the time offset on the classification performance.

\begin{figure*}
\centering
\subfigure[8-PSK]{%
\includegraphics[width=0.35\textwidth,height=!]{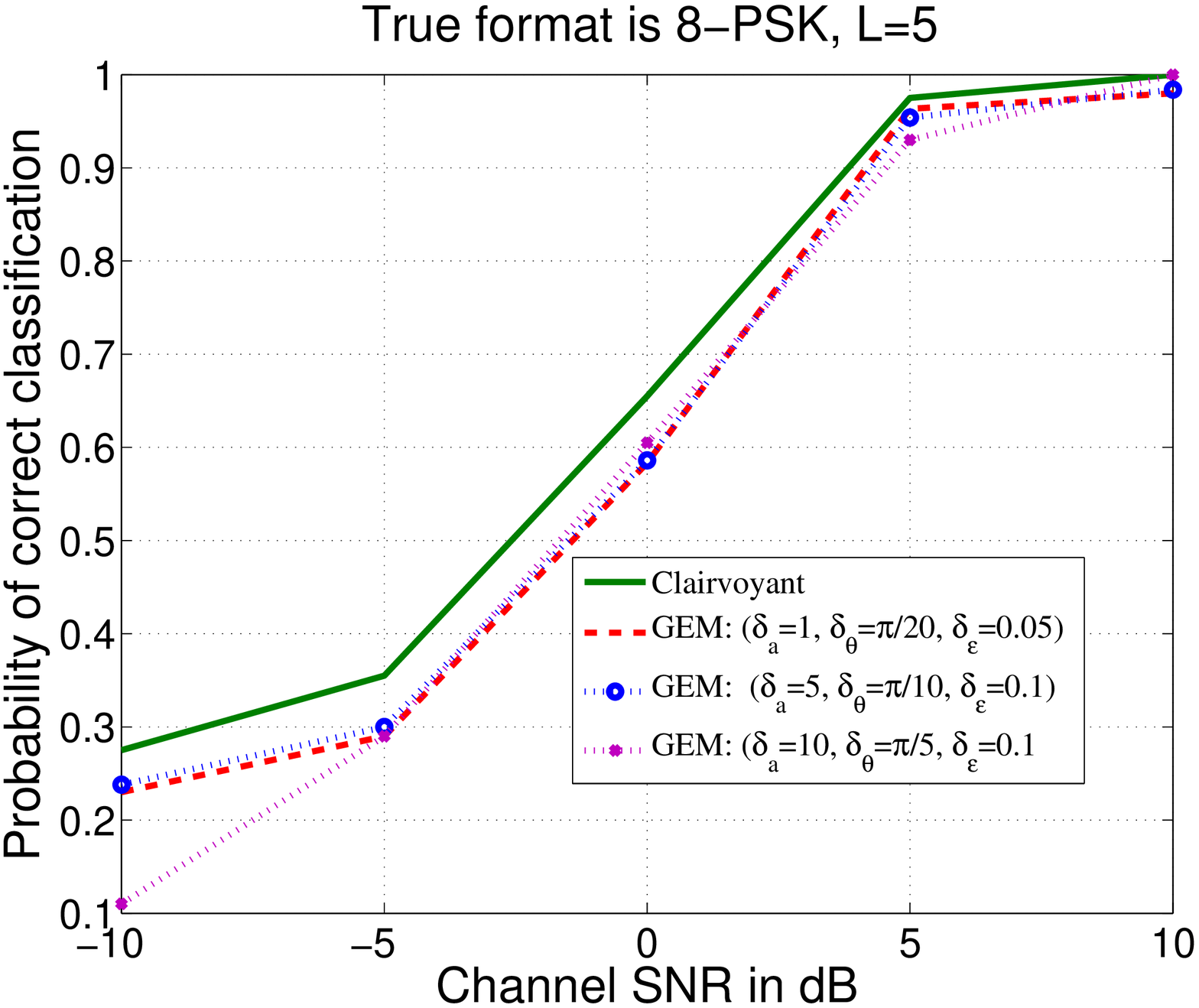}
}
\quad
\subfigure[8-QAM]{%
\includegraphics[width=0.35\textwidth,height=!]{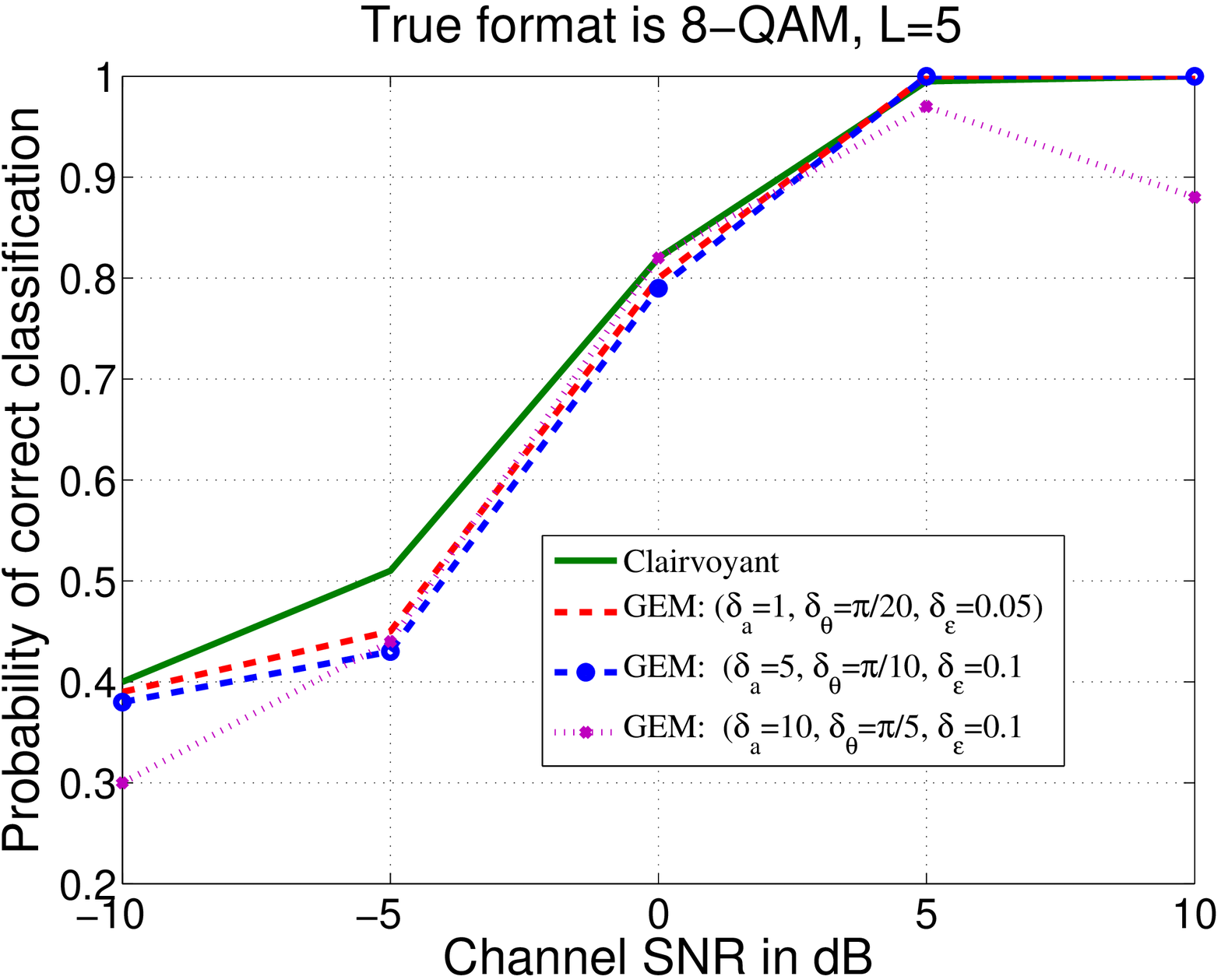}
}
\subfigure[16-PSK]{%
\includegraphics[width=0.35\textwidth,height=!]{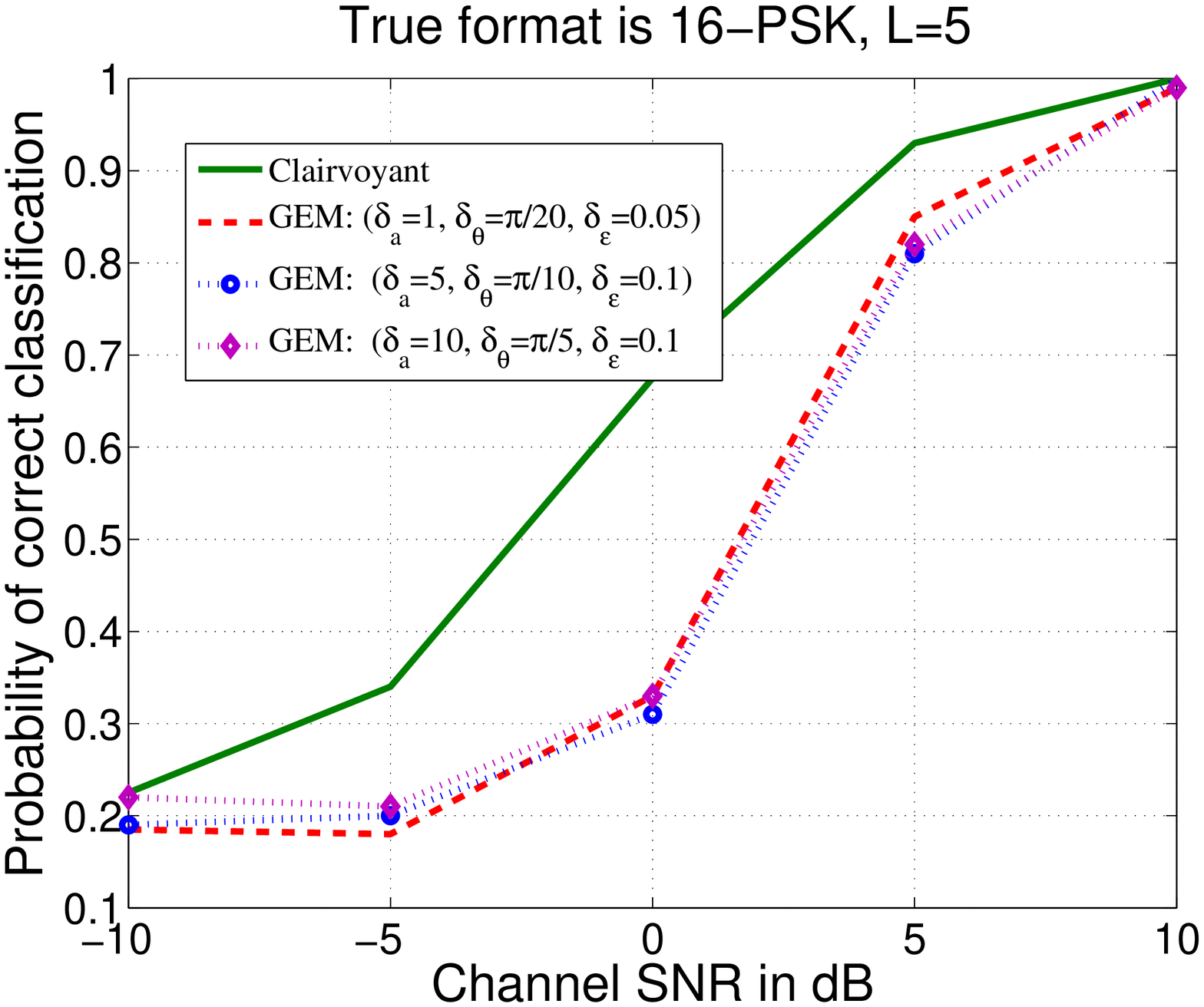}
}
\quad
\subfigure[16-QAM]{%
\includegraphics[width=0.35\textwidth,height=!]{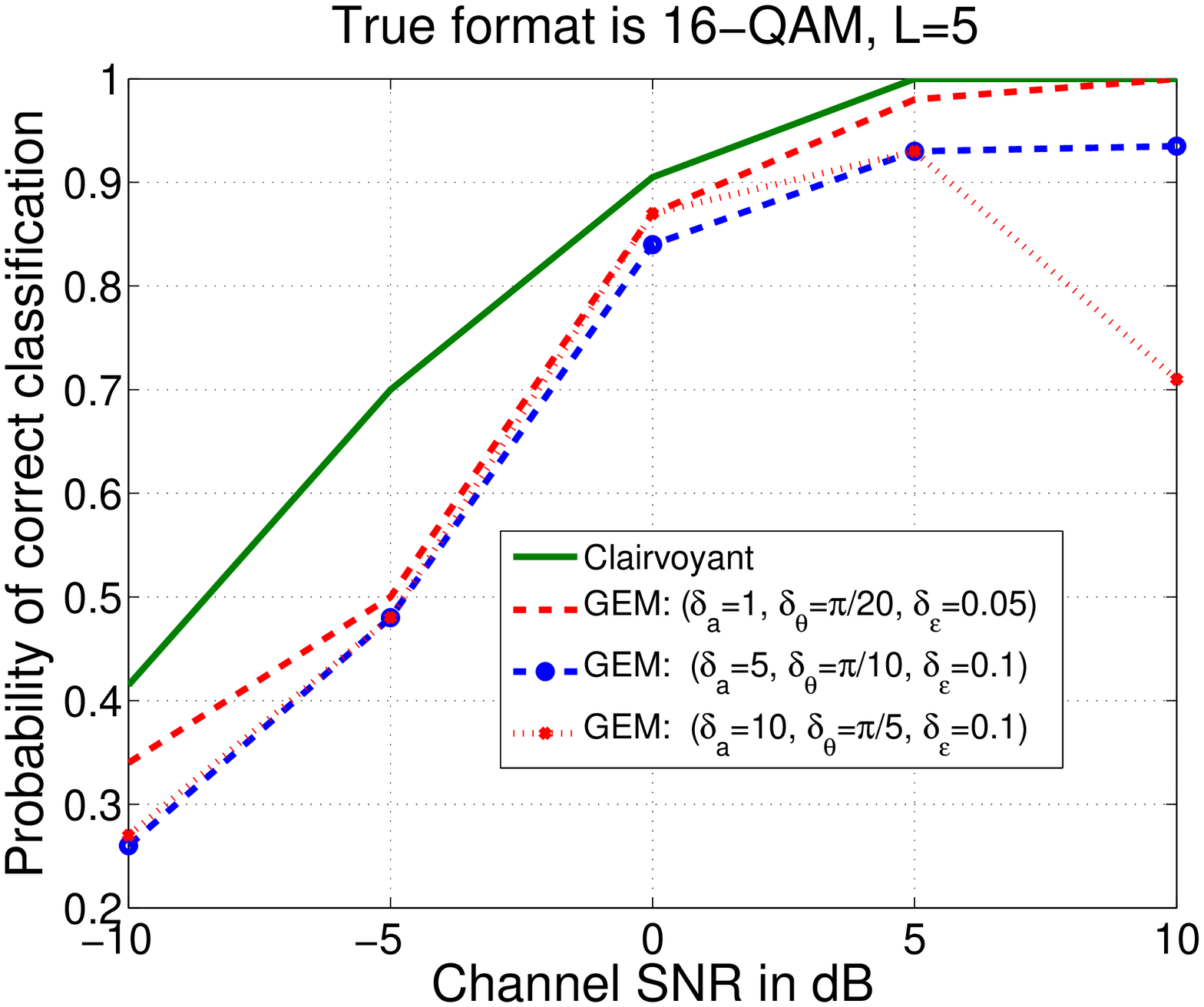}
}
%
\caption{Impact of the initial values of unknowns on  the GEM performance;  $L=5$, $N=100$, initial values for GEM are taken based on  $(\delta_{a}=1$, $\delta_{\theta}=\pi/20$,  $ \delta_{\varepsilon}=0.05$), $(\delta_{a}=5$, $\delta_{\theta}=\pi/10$,  $ \delta_{\varepsilon}=0.1)$, and $(\delta_{a}=10$, $\delta_{\theta}=\pi/5$,  $ \delta_{\varepsilon}=0.1)$}
\label{fig_initial}
\end{figure*}

\subsection{Impact of initialization of unknowns on the GEM performance}\label{initial}
It is  known that the performance of the GEM algorithm is highly sensitive to initialization of the unknown parameters. When the initial values deviate  significantly   from the true estimates, it becomes highly likely that the estimates of GEM are trapped in local maxima especially when the number of unknowns is large.
 To demonstrate the impact of the initialization on the performance of the GEM algorithm, we take initialization points of unknown parameters  as the true values plus some error. More specifically, we consider different scenarios where the initial values for the unknown parameters $a_l$, $\theta_l$ and $\varepsilon_l$ can take any random values uniformly distributed in the regions  $[0, a_{l}+\delta_{a}]$, $[\theta_l-\delta_{\theta}, ~\theta_l+\delta_{\theta}]$,   and $[\varepsilon_l-\delta_{\varepsilon}, ~\varepsilon_l+\delta_{\varepsilon}]$, respectively, for $l=1,\cdots,L$ where  $\delta_{a} , \delta_{\theta} ,\delta_{\varepsilon} > 0$ are the maximum errors for each unknown. These error bounds determine how close the initial points are to the true values.

In Fig. \ref{fig_initial}, we plot the probability of correct classification   vs channel SNR. Given the $i$-th modulation format, the probability of correct classification is denoted by $P(\mathcal H_i|\mathcal H_i)$ which means that the classifier decides $\mathcal H_i$ when the true hypothesis is $\mathcal H_i$.   We let $L=5$, $N=100$. Three sets of initial values are considered taking  $\{\delta_{a}=1, \delta_{\theta}=\pi/20,   \delta_{\varepsilon}=0.05\}$, $\{\delta_{a}=5, \delta_{\theta}=\pi/10,  \delta_{\varepsilon}=0.1\}$, and $\{\delta_{a}=10, \delta_{\theta}=\pi/5,  \delta_{\varepsilon}=0.1\}$.
Each result is based on 500 Monte Carlo runs.   We also plot the probability of correct classification with Clairvoyant classifier which assumes that the unknown parameter vector $\mathbf u$ is known and the classification is carried out based on the marginalized likelihood function over the constellation symbols.

With the first two sets of values for error,  the initial values of unknowns  are  not considerably  away from the true values. In this case, it can be seen from Fig. \ref{fig_initial} that the GEM algorithm  is capable of providing performance that is comparable to  the Clairvoyant classifier  when the  true modulation format consists of small constellations. When 16-PSK or 16-QAM is the true format, there is a certain performance gap between GEM based and Clairvoyant classifier in the low SNR region. With the third  set of values for error, the initial values can be considerably  away from the true values of unknowns.  Based on Fig. \ref{fig_initial}, when the true modulation format is either 8-PSK, or 16-PSK the GEM algorithm does not seem to depend much on the initial values even though they (initial values) considerably deviate from the actual values.  However, when  8-QAM or 16-QAM is  the  true format, it is observed that the GEM performance seems to degrade as the   initial values significantly deviate from the true values, indicating that the classification of QAM modulations is more sensitive to initialization. In the case of multiple sensors (where the number of unknowns is proportional to the number of sensors), the likelihood function may exhibit a large number of local maxima. Therefore, when the initial values are significantly far away from the true estimates, the GEM estimates for unknowns  can be easily trapped at  local maxima leading to poor performance.


 \subsubsection{Number of sensors}
\begin{figure}[ht]
\centering
\includegraphics[width=0.45\textwidth,height=!]{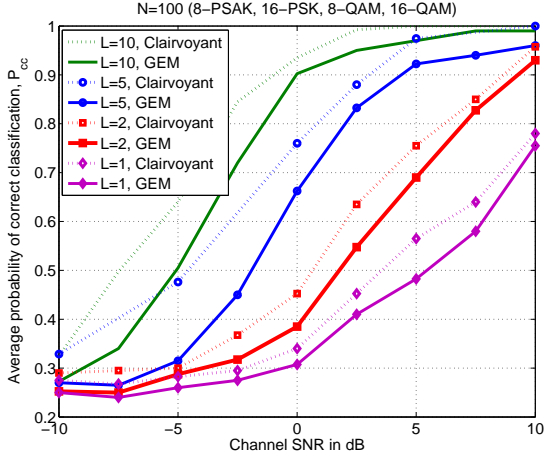}
\caption{Performance of AMC with GEM  as $L$ varies: Initialization for GEM  is obtained from    $\delta_{a}=5$, $\delta_{\theta}=\pi/10$, and $ \delta_{\varepsilon}=0.1$, $N=100$}
\label{fig_L}
\end{figure}

 In Fig. \ref{fig_L}, we plot the average (taken over all modulation formats) probability of correct classification versus channel SNR as the number of sensors varies. The average probability of correct classification $P_{cc}$ is defined as $P_{cc}\triangleq 1/S\sum_{i=1}^S P(\mathcal H_i|\mathcal H_i)$.   The results are based on the GEM algorithm with  $\delta_{a}=5$, $\delta_{\theta}=\pi/10$, and $ \delta_{\varepsilon}=0.1$.  In  Fig.  \ref{fig_L}, it can be seen that  there is a significant  performance improvement as the number of sensors increases.  For example, when SNR is $5dB$,  AMC with negligible classification error using GEM can be achieved employing $10$ sensors whereas  the average $P_{cc}$ is  below $0.5$ with a single sensor. Furthermore, when $L$ is small, it is observed that the GEM algorithm  provides performance that is comparable to the  Clairvoyant classifier. When $L$ is increased, the performance gap between GEM based and  Clairvoyant classifiers also increases, especially in the low SNR region.  As the SNR  increases, however,  the performance gap between two classifiers is not significant  irrespective of $L$.

\begin{figure}[h]
\begin{center}
\includegraphics[width=0.45\textwidth,height=!]{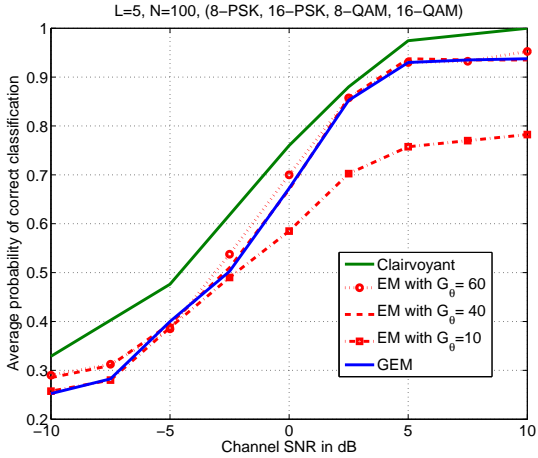}
\end{center}
\vspace{-.2in}
\caption{\small Performance of AMC with   EM and  GEM algorithms: $\delta_{a}=5$, $\delta_{\theta}=\pi/10$, and $ \delta_{\varepsilon}=0.1$,   $N=100$, $L=5$}\label{fig_initial_EM_GEM}
\end{figure}

\begin{figure}[h]
\begin{center}
\includegraphics[width=0.45\textwidth,height=!]{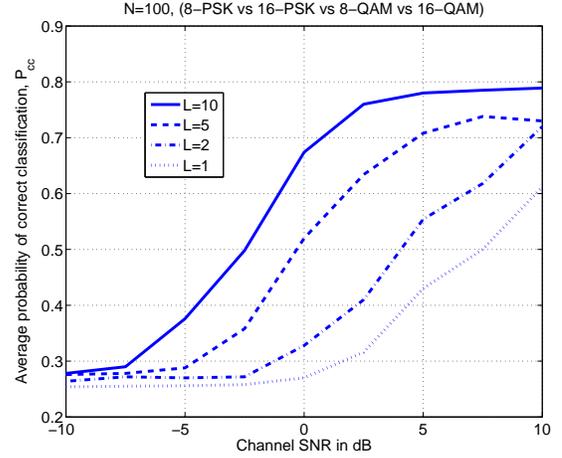}
\end{center}
\vspace{-.2in}
\caption{\small  Performance of  GEM with SA  based initialization  (in Algorithm \ref{SA});  $N=100$}\label{fig:pc}
\end{figure}

\subsubsection{EM and GEM}
The main motivation for us to use GEM instead of the EM algorithm is the high computational complexity associated in the maximization step (\ref{mle4}) of the EM algorithm.  In Fig. \ref{fig_initial_EM_GEM} we compare the performance  of AMC with both the EM and GEM algorithms. We use average probability of correct classification $P_{cc}$ as the performance criterion  in Fig. \ref{fig_initial_EM_GEM}. To implement the maximization step  of EM in (\ref{mle4}), we perform a two dimensional grid search over $\epsilon_l$ and $\theta_l$. It is  noted that, with the GEM algorithm, a line search method is employed  to estimate $\epsilon_l$ based on (\ref{mle6}). In Fig. \ref{fig_initial_EM_GEM}, we fix the number of grid points along $\epsilon_l$ (at $50$) and vary the number of grid points over $\theta_l$ (denoted by $G_{\theta}$) for EM. We use the same initialization values for unknowns for both the algorithms and set $\{\delta_{a}=5, \delta_{\theta}=\pi/10,  \delta_{\varepsilon}=0.1\}$. It is observed that, with  a fine  grid for 2-dimensional optimization in the EM algorithm,  there is negligible  performance loss when  using GEM instead of  EM in terms of $P_{cc}$. However, it should be noted that the finer the grid, the higher the  computational complexity of EM.  With a coarse grid ($G_{\theta} = 10$), it is observed that  EM performs worse  than  GEM. Thus, GEM, which requires only 1-dimensional grid search,  appears to be a better  choice over EM. It is worth mentioning  that, we have used   naive approaches   for 1-D and 2-D optimization problems (line search  and 2-D grid search) to provide a fair  comparison.

\begin{table}
\caption{Ratio between Run times required for EM and GEM algorithms; $SNR=0dB$}
\centering
\centering
\begin{small}
\begin{tabular}{|l|l|l|l|}
  \hline
 $~$&$L=1$ & $L=2$ & $L=5$ \\
  \hline
   Run time ratio EM/GEM ($G_{\theta}=40$) &$1.31$ & 1.32 & $1.36$ \\
   \hline
  Run time ratio EM/GEM ($G_{\theta}=60$) & $1.54$& $1.50$ &  $1.53$ \\
     \hline
 \end{tabular}
\end{small}\label{table_runtime}
\end{table}

In Table \ref{table_runtime}, we compare the computational complexity in terms of the average  run time required to make the classification decision based on GEM and and EM algorithms. We obtained the average run time over $500$ Monte Carlo runs in MATLAB R2013a with 64-bit operating
system in Intel(R) Core(TM) i7-3770 CPU$@$ 3.40 GHz.  We provide the ratio between run times required by EM and GEM.   Again, we fixed the grid points over $\epsilon_l$ for EM and GEM at the same value and $G_{\theta}$ is varied. We further  use the same initialization points for both algorithms  and let $L=5$ and $SNR=0dB$. From Table \ref{table_runtime}, it can be seen for this particular scenario that the computational complexity of EM with a fine grid  is approximately  $1.5$ times  that of GEM on average, to attain  the same $P_{cc}$ performance.

\subsection{Performance of GEM with SA based initialization}
Next, we investigate the performance of the GEM algorithm considering  SA,  stated in Algorithm \ref{SA}, as the initialization technique. Using SA, we first get a rough estimate for the initial values of the unknowns using a coarse grid. The accuracy depends on the grid size. As the grid becomes finer, the complexity of the algorithm  increases. For SA based initialization as stated in Algorithm \ref{SA}, we set $a_l^U=F_A^{-1}(0.99;\sigma)$ for $l=1,\cdots, L$, where $F_A^{-1}(\cdot;\sigma)$ is the inverse cumulative distribution function of Rayleigh distribution parameterized by $\sigma$. The grid increments for initialization ($\Delta$s) are selected such that we have $10$ grid points for each unknown, i.e., $\Omega$ consists of $1000$ points. We set $K=200$ and $d=1.6$ for the SA algorithm. Note that our SA algorithm requires only $200$ evaluations of the likelihood function for initialization as opposed to $1000$ that would be required by an exhaustive grid search.

In Fig. \ref{fig:pc},  we plot the performance of the GEM classifier in terms of  $P_{cc}$  with different number of sensors when the initial values are selected based on the SA algorithm.
It is observed that the performance of the GEM with  SA based initialization is monotonically increasing  in the low-to-moderate SNR region for all $L$. Thus, it appears that GEM with SA based initialization scheme  is a promising technique for AMC  with any  given number of sensors.

\begin{figure}[h]
\begin{center}
\includegraphics[width=0.45\textwidth,height=!]{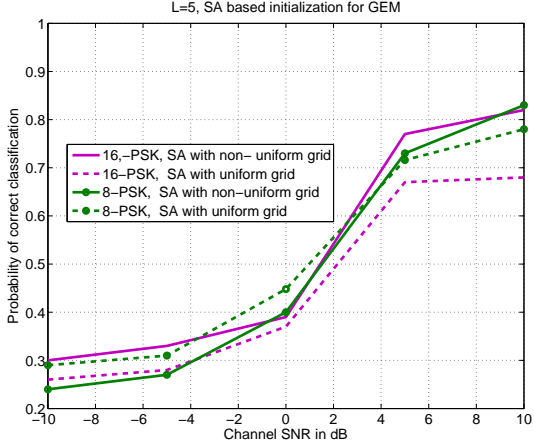}
\end{center}
\vspace{-.2in}
\caption{\small  Performance of the GEM algorithm: SA based initialization with different grid sizes:   $N=100$, $L=5$}\label{fig:pc_d_nonuniform}
\end{figure}
It is noted that when the initial values are not significantly  far away from the true values of unknowns as shown  in Fig. \ref{fig_L}, a considerable performance gain is achieved with multiple sensors  compared  to a single sensor. Comparing Figs. \ref{fig_L} and  \ref{fig:pc}, it is seen that although GEM with SA based initialization provides acceptable performance,  there is room for further improvement.  While it is expected that the performance of the GEM algorithm with  SA based initialization could be further improved by increasing the number of  grid points, it is not desirable due to higher computational complexity at the initialization stage. To solve this problem  to a certain extent, we created a nonuniform grid for SA based initialization, where the number of grid points along $\theta_l$ are increased while those along $a_l$ are reduced, so that the total number of  grid points in $\Omega$ are kept the same compared to that in the uniform grid considered above. The motivation behind the use of a nonuniform grid is the observation that  it is the channel phase that is incorrectly estimated most of the time  with a uniform grid. In Fig. \ref{fig:pc_d_nonuniform},  we plot the probability of correct classification for $L=5$  when the true format is either 8-PSK or 16-PSK. For the nonuniform grid, we take  $5$, $20$ and $10$ grid points for $a_1$, $\theta_l$ and $\varepsilon_l$, respectively.  From Fig. \ref{fig:pc_d_nonuniform}, we observe an improved  performance of GEM with SA based initialization with nonuniform grid compared to a uniform grid with 16-PSK.  With 8-PSK, the performance with nonuniform grid is better than that with the uniform grid only when the SNR is higher. Further, while curves are not included, the performance when the true format is either 8-QAM or 16-QAM does not seem to vary significantly with a nonuniform grid compared to a uniform grid.

\begin{figure}[h]
\begin{center}
\includegraphics[width=0.45\textwidth,height=!]{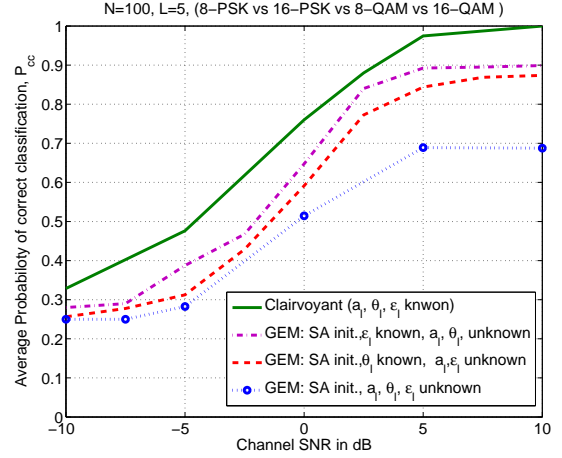}
\end{center}
\vspace{-.2in}
\caption{\small  Performance of the GEM algorithm  as the number of unknowns varies; average probability of classification vs SNR; $L=5$, $N=100$}\label{fig:pc_d_unknowns}
\end{figure}


Next, we investigate  the effectiveness of the SA based initialization scheme for the  GEM algorithm   with multiple sensors  by varying the number of unknowns. It is noted that, we consider three unknown parameters  for  each  node (i.e. $a_l$, $\theta_l$ and $\varepsilon_l$ at the $l$-th node for $l=1, \cdots, L$). In Fig.  \ref{fig:pc_d_unknowns},  we plot the performance in terms of  $P_{cc}$ for $L=5$ as the number of unknowns is varied.    It can be observed that, if either the time offset  $\epsilon_l$ or the  channel phase $\theta_l$ at the $l$-th sensor is  assumed to be known, then  the GEM algorithm with SA based initialization (with a uniform grid) provides performance that is closer   to the  Clairvoyant classifier. In particular, if  fewer number of parameters per node have  to be estimated via ML estimation, then  GEM with SA as the initialization technique with a coarse grid provides acceptable  performance compared to the  Clairvoyant classifier.

%

From Figs.  \ref{fig_initial}-\ref{fig:pc_d_unknowns}, we conclude the following:    (i). Given a relatively   good initialization technique, GEM for HML based AMC for linear modulation classification   is capable of providing promising performance as the number of sensors  increases. In depth investigation of initialization schemes for GEM is beyond the scope of this paper. (ii). When  SA with a coarse grid is chosen as the initialization scheme, GEM provides good performance, especially  in the  low-to moderate  SNR region considered in this  paper. When the  number of unknowns per sensor is small, GEM with SA based initialization provides performance that is closer to the Clairvoyant classifier.

In the following, we further investigate the performance of the  GEM algorithm for AMC with multiple sensors with respect to several other parameters.

\subsection{Number of samples per node}
\begin{figure}[h]
\begin{center}
\includegraphics[width=0.45\textwidth,height=!]{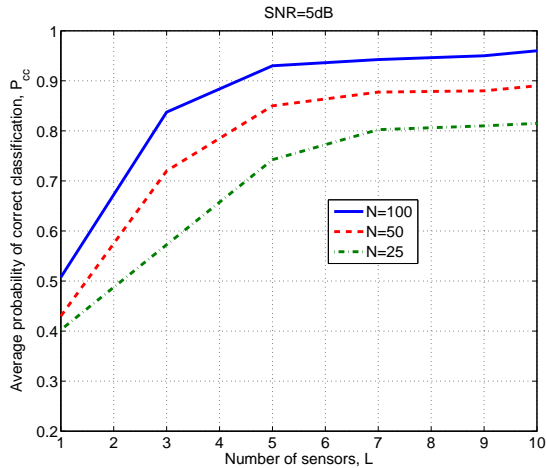}
\end{center}
\vspace{-.2in}
\caption{Average probability of correct classification $P_{cc}$  vs number of sensors $L$ as the number of samples per node $N$ varies, $SNR=5dB$,  initial estimates for unknown parameters for GEM are taken as described in the first paragraph of  subsection \ref{initial} with errors $\delta_{a}=5$, $\delta_{\theta}=\pi/10$, and $ \delta_{\varepsilon}=0.1$ }\label{fig_samples}
\end{figure}

Next,  we illustrate the impact of the number of samples per node on the classification performance as the number of sensors varies. In Fig. \ref{fig_samples}, the  average probability of correct classification vs the number of sensors is plotted as the number of samples per node, $N$, varies when $SNR=5dB$. In Fig. \ref{fig_samples}, GEM is performed with the initialization scheme as considered in Section \ref{initial} with $\delta_{a}=5$, $\delta_{\theta}=\pi/10$, and $ \delta_{\varepsilon}=0.1$.  Results in Fig. \ref{fig_samples} validate  the claim that a  relatively  small number of samples  at each node is capable of providing  a closer  performance based on the proposed approach as the number of sensors increases even if the SNR is low.

\subsection{Performance with other comparable classifiers}
\begin{figure}[ht]
\centering
\subfigure[L=1, N=100]{%
\includegraphics[width=0.55\textwidth,height=!]{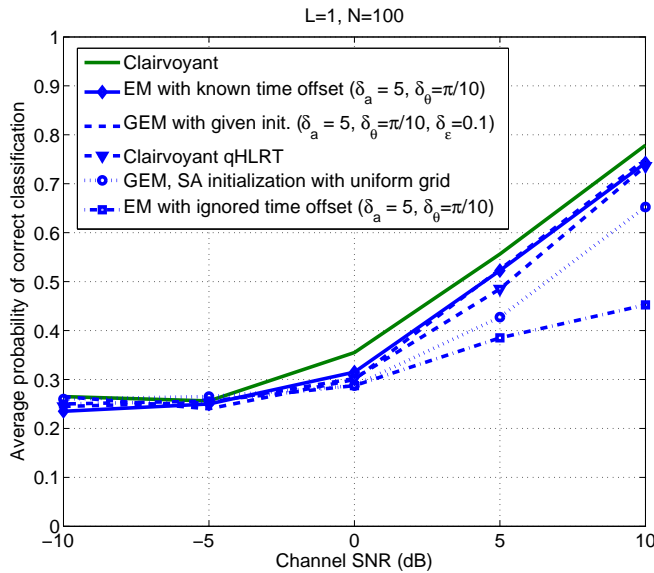}
}
\quad
\subfigure[L=5, N=100]{%
\includegraphics[width=0.55\textwidth,height=!]{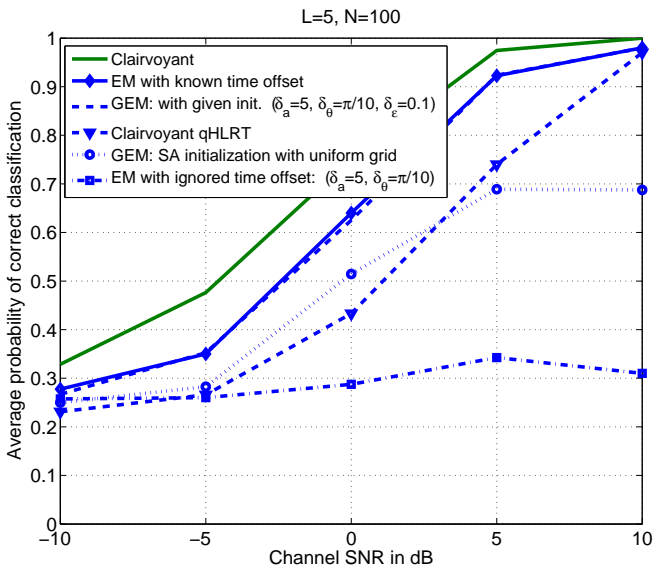}
}
\caption{Average probability of correct classification $P_{cc}$ vs. channel SNR: $N=100$, for $L=1$, GEM is implemented with SA as the initialization scheme. For $L=5$, GEM is performed when the initialization points are selected based on $\{\delta_a = 5, \delta_\theta = \pi/10, \delta_{\epsilon} = 0.1\}$ as well as based on SA with a  uniform grid }
\label{fig:avgpc}
\end{figure}

In Fig.   \ref{fig:avgpc}, we compare  the proposed GEM based classifier with four other classifiers: 1) Clairvoyant classifier of \cite{mendel_tc_00} which has perfect information on $\mathbf{u}$, 2) Clairvoyant EM classifier which has perfect information on $\varepsilon_l$, but does not know $a_l$ or $\theta_l$, for $l=1,\cdots,L$, 3) the qHLRT based multi-antenna classifier proposed in \cite{inkol_tc14}, which has perfect information on $\varepsilon_l$, but does not know $a_l$ or $\theta_l$, for $l=1,\cdots,L$, and 4) EM classifier which ignores time offsets, i.e., which assumes that time offsets are zero. The classifier proposed in \cite{inkol_tc14} uses moment-based estimators and combines the log-likelihood values (from sensors/antennas) using a weighted average to perform classification. In order to provide a fair comparison, we replaced the noise variance estimates in \cite{inkol_tc14} with true noise variance in qHLRT. We call this classifier the Clairvoyant qHLRT due to the fact that it has perfect information on $\varepsilon_l$ for $l=1,\cdots,L$. In Fig. \ref{fig:avgpc} (a), we let $L=1$, $N=100$ while in Fig. \ref{fig:avgpc} (b), we let $L=5$ and $N=100$. We consider two initialization  schemes for GEM in Fig. \ref{fig:avgpc}. The  dashed curve is the average $P_{cc}$ when the initial values are selected as true value plus some error as considered in Section \ref{initial} where $\{\delta_a = 5, \delta_\theta = \pi/10, \delta_{\epsilon} = 0.1\}$. The dotted curve with circle markers is for GEM with SA based initialization.   It is clear from Figs. \ref{fig:avgpc} (a) and \ref{fig:avgpc} (b), that when time offsets are ignored, the performance of the resulting classifier is extremely poor. This result indicates the fact that residual time offsets need to be taken into account in a modulation classification application. We can see from Fig. \ref{fig:avgpc} (b) that the performance of the proposed GEM based modulation classifier with a relatively  good initialization scheme  is almost identical  to the Clairvoyant EM classifier that has perfect information on $\varepsilon_l$ for $l=1,\cdots, L$. Further,  Figs.  \ref{fig:avgpc} (a) and \ref{fig:avgpc} (b) again verify  that,  with a good initialization scheme for GEM,  a significant performance gain can be obtained by increasing the number of sensors  compared to that with a single sensor. With SA based initialization, the performance gap between GEM and Clairvoyant classifiers is smaller for  single sensor than that with $L=5$.
The Clairvoyant classifier \cite{mendel_tc_00} serves as an upper bound on the performance of all modulation classifiers. It is also interesting to see that the Clairvoyant EM classifier performs very close to this upper bound even though it knows neither the channel phase nor the channel gain. The Clairvoyant EM is also superior to the Clairvoyant qHLRT even though both have perfect information on time offsets. This result is due to the fact that moment-based estimators used in \cite{inkol_tc14} are suboptimal (in terms of maximizing the likelihood function) and they do not take into account coupling between different sensor observations due to common unknown constellation symbols.  The trade-off, however, is the computational complexity associated with EM, i.e., EM is an iterative algorithm with higher computational complexity than qHLRT which is based on single shot moment-based estimates. Furthermore, as seen in Figs. \ref{fig:avgpc} (a) and \ref{fig:avgpc} (b),  the proposed GEM algorithm  (without the knowledge of channel gain, channel phase and the time offset)  with a good initialization scheme  also provides   performance that is very close  to the Clairvoyant classifier.

%
%
%
%

\section{Conclusion}\label{sec:conc}
We have addressed  the problem of linear modulation classification with multiple sensors in the presence of unknown time offset in addition to unknown phase offset and received signal amplitude. We considered a centralized fusion scheme, where multiple sensors transmit their observations to a central fusion center to perform  classification.  We have proposed a novel hybrid maximum likelihood (HML) approach where the unknowns are estimated using a tractable GEM algorithm. We have shown  that the performance of AMC can be significantly improved as the number of sensors increases  when  a good initialization technique for GEM is employed. Our proposed approach employs only a small number of samples to perform both time/phase synchronization and modulation classification. The simulation results show that the proposed approach provides excellent classification performance with only a small number of samples.

In this paper, we assumed  that the sensors transmit their raw observations to a fusion center to perform classification. An interesting future avenue is to consider the AMC problem when the sensors transmit only a summary of the observations to a fusion center. Further, AMC considering   delay jitter instead of a  fixed delay for  all the observation symbols is an another interesting future direction.

\bibliographystyle{IEEEtran}
\bibliography{Journal,Conf,Book}

\begin{thebibliography}{10}
\providecommand{\url}[1]{#1}
\csname url@samestyle\endcsname
\providecommand{\newblock}{\relax}
\providecommand{\bibinfo}[2]{#2}
\providecommand{\BIBentrySTDinterwordspacing}{\spaceskip=0pt\relax}
\providecommand{\BIBentryALTinterwordstretchfactor}{4}
\providecommand{\BIBentryALTinterwordspacing}{\spaceskip=\fontdimen2\font plus
\BIBentryALTinterwordstretchfactor\fontdimen3\font minus
  \fontdimen4\font\relax}
\providecommand{\BIBforeignlanguage}[2]{{%
\expandafter\ifx\csname l@#1\endcsname\relax
\typeout{** WARNING: IEEEtran.bst: No hyphenation pattern has been}%
\typeout{** loaded for the language `#1'. Using the pattern for}%
\typeout{** the default language instead.}%
\else
\language=\csname l@#1\endcsname
\fi
#2}}
\providecommand{\BIBdecl}{\relax}
\BIBdecl

\bibitem{hamkins_06}
J.~Hamkins, M.~K. Simon, and J.~H. Yuhen, \emph{Autonomous Software-Defined
  Radio Receivers for Deep Space Applications (JPL Deep-Space Communications
  and Navigation Series)}.\hskip 1em plus 0.5em minus 0.4em\relax
  Wiley-Interscience, 2006.

\bibitem{su_iet07}
O.~A. Dobre, A.~Abdi, Y.~Bar-{Ness}, and W.~Su, ``Survey of automatic
  modulation classification techniques: classical approaches and new trends,''
  \emph{IET Communications}, vol.~1, no.~2, pp. 137--159, Apr. 2007.

\bibitem{su_tsmc_11}
J.~L. Xu, W.~Su, and M.~Zhou, ``Likelihood-ratio approaches to automatic
  modulation classification,'' \emph{IEEE Trans. Systems, Man, and Cybernetics
  - Part C: Applications and Reviews}, vol.~41, no.~4, pp. 455--469, Jul. 2011.

\bibitem{poor_det}
H.~V. Poor, \emph{An introduction to detection and estimation}.\hskip 1em plus
  0.5em minus 0.4em\relax Springer, 1994.

\bibitem{su_sarnoff05}
O.~A. Dobre, A.~Abdi, Y.~Bar-{Ness}, and W.~Su, ``Blind modulation
  classification: a concept whose time has come,'' in \emph{Proc. IEEE Sarnoff
  Symposium on Advances in Wired and Wireless Comm.}, Apr. 2005.

\bibitem{dobre_twc09}
F.~Hameed, O.~A. Dobre, and D.~C. Popescu, ``On the likelihood-based approach
  to modulation classification,'' \emph{IEEE Trans. Wireless Comm.}, vol.~8,
  no.~12, pp. 5884--5892, Dec. 2009.

\bibitem{silva_tc11}
V.~G. Chavali and C.~R. C.~M. {da Silva}, ``Maximum-likelihood classification
  of digital amplitude-phase modulated signals in flat fading non-{G}aussian
  channels,'' \emph{IEEE Trans. Commun.}, vol.~59, no.~8, pp. 2051--2056, Aug.
  2011.

\bibitem{mengali_97}
U.~Mengali and A.~N. {D'A}ndrea, \emph{Synchronization Techniques for Digital
  Receivers}.\hskip 1em plus 0.5em minus 0.4em\relax New York: Plenum, 1997.

\bibitem{weber_tc98}
B.~F. Beidas and C.~L. Weber, ``Asynchronous classification of {MFSK} signals
  using the higher order correlation domain,'' \emph{IEEE Trans. Commun.},
  vol.~46, no.~4, pp. 480--493, Apr. 1998.

\bibitem{namazi_tc02}
A.~E. {El-Mahdy} and N.~M. Namazi, ``Classification of multiple {M-ary}
  frequency-shift keying signals over a {R}ayleigh fading channel,'' \emph{IEEE
  Trans. Commun.}, vol.~50, no.~6, pp. 967--974, Jun. 2002.

\bibitem{polydoros_95}
C.~Huang and A.~Polydoros, ``Likelihood methods for {MPSK} modulation
  classification,'' \emph{IEEE Trans. Commun.}, vol.~43, no. 234, pp.
  1493--1504, 1995.

\bibitem{guan_wcnc08}
Q.~Shi, Y.~Gong, and L.~Guan, ``Asynchronous classification of high-order
  {QAM}s,'' in \emph{Proc. IEEE Wireless Comm. and Networking Conf. (WCNC)},
  Apr. 2008.

\bibitem{silva_tc11_2}
W.~C. Headley and C.~R. C.~M. {da Silva}, ``Asynchronous classification of
  digital amplitude-phase modulations in flat-fading channels,'' \emph{IEEE
  Trans. Commun.}, vol.~59, no.~1, pp. 7--12, Jan. 2011.

\bibitem{ozdemir_globecom13}
O.~Ozdemir, P.~K. Varshney, and W.~Su, ``Asynchronous hybrid maximum likelihood
  classification of linear modulations,'' in \emph{Globecom}, Dec. 2013.

\bibitem{mendel_tc_00}
W.~Wei and J.~M. Mendel, ``Maximum-likelihood classification for digital
  amplitude-phase modulations,'' \emph{IEEE Trans. Commun.}, vol.~48, no.~2,
  pp. 189--193, Feb. 2000.

\bibitem{sadler_tc00}
A.~Swami and B.~M. Sadler, ``Hierarchical digital modulation classification
  using cumulants,'' \emph{IEEE Trans. Commun.}, vol.~38, no.~3, pp. 416--429,
  Mar. 2000.

\bibitem{cabric_cl11}
P.~Urrizo, E.~Rebeiz, P.~Pawelczak, and D.~Cabric, ``Computationally efficient
  modulation level classification based on probability distribution distance
  functions,'' \emph{IEEE Commun. Letters}, vol.~15, no.~5, pp. 476--478, May
  2011.

\bibitem{rubin_jrs77}
A.~P. Dempster, N.~M. Laird, and D.~B. Rubin, ``Maximum likelihood from
  incomplete data via the {EM} algorithm,'' \emph{Journal Roy. Stat. Soc.
  (Series B)}, vol.~39, no.~1, pp. 1--38, 1977.

\bibitem{wu_ann83}
C.~F.~J. Wu, ``On the convergence properties of the {EM} algorithm,''
  \emph{Ann. Stat.}, vol.~11, no.~1, pp. 95--103, 1983.

\bibitem{Proakis:book}
J.~Proakis, \emph{Digital Communications}.\hskip 1em plus 0.5em minus
  0.4em\relax McGraw Hill, 1995.

\bibitem{inkol_tc14}
A.~R.-Kebrya, I.-M. Kim, D.~I. Kim, F.~Chan, and R.~Inkol, ``Likelihood-based
  modulation classification for multiple-antenna receiver,'' \emph{IEEE Trans.
  Commun.}, vol.~61, no.~9, pp. 3816--3829, 2013.

\end{thebibliography}
\end{document}